\documentclass[referee]{raa}           

\usepackage{graphicx,times}
\usepackage{natbib}
\usepackage{amssymb,amsmath}
\bibpunct{(}{)}{;}{a}{}{,}

\usepackage[pagebackref=true]{hyperref}

\begin{document}

   \title{Photometric metallicities of 0.8 million KiDS stars}

 \volnopage{ {\bf 20XX} Vol.\ {\bf X} No. {\bf XX}, 000--000}
   \setcounter{page}{1}

   \author{Bao-Kun Sun
   \inst{1}, Bing-Qiu Chen\inst{1}, Xiao-Wei Liu\inst{1}
   }

   \institute{$^1$South-Western Institute for Astronomy Research, Yunnan University, Chenggong District, Kunming 650091, P.\,R. China; {\it bchen@ynu.edu.cn; x.liu@ynu.edu.cn}\\
\vs \no
   {\small Received 20XX Month Day; accepted 20XX Month Day}
}

\abstract{Accurate determinations of metallicity for large, complete stellar samples are essential for advancing various studies of the Milky Way. In this paper, we present a data-driven algorithm that leverages photometric data from the KiDS and the VIKING surveys to estimate stellar absolute magnitude, effective temperature and metallicities. The algorithm is trained and validated using spectroscopic data from LAMOST, SEGUE, APOGEE, and GALAH, as well as a catalog of very metal-poor stars from the literature, and Gaia EDR3 data. This approach enables us to estimate metallicities, effective temperatures, and $g$-band absolute magnitudes for approximately 0.8 million stars in the KiDS dataset. The photometric metallicity estimates exhibit an uncertainty of around 0.28\,dex when compared to spectroscopic studies, within the metallicity range of $-$2 dex to 0.5\,dex. The photometric effective temperature estimates have an uncertainty of around 149\,K, while the uncertainty in the absolute magnitude is approximately 0.36\,mag. The metallicity estimates are reliable for values down to about $-$2\,dex. This catalog represents a valuable resource for studying the structure and chemical properties of the Milky Way, offering an extensive dataset for future investigations into Galactic formation and evolution.
}

   \authorrunning{B.-K. Sun et al. }            
   \titlerunning{Metallicities of KiDS stars}  
   \maketitle

%
\section{Introduction}           
\label{sect:intro}
Understanding the chemical composition of the Milky Way is key to unraveling its structure, formation, and evolutionary processes \citep[e.g.,][]{casagrande2011new, peng2013stellar, yan2019chemical, whitten2021photometric, martin2023pristine, hackshaw2024x, sun2023spar}.

The field has been transformed by large-scale spectroscopic surveys, such as the Radial Velocity Experiment (RAVE; \citealt{steinmetz2006radial}), Sloan Extension for Galactic Understanding and Exploration (SEGUE; \citealt{yanny2009segue}), Large Sky Area Multi-Object Fiber Spectroscopic Telescope (LAMOST; \citealt{zhao2012lamost, liu2015preface}), Apache Point Observatory Galactic Evolution Experiment (APOGEE; \citealt{majewski2017apache}), and the Galactic Archaeology with HERMES project (GALAH; \citealt{de2015galah}). These surveys have collected spectra for more than ten million stars \citep[e.g.,][] {luo2015first, yuan2015lamost}, providing critical stellar parameters such as effective temperature ($T_{\rm eff}$), metallicity ([Fe/H]), surface gravity ($\log g$), and other physical properties. This wealth of data has enabled significant advances in our understanding of the Milky Way's structure, chemical composition, and kinematics.

For example, using metallicity data from LAMOST, \citet{wang2019metallicity} found that the vertical metallicity gradient is steeper in younger stars, reaches a maximum in stars of intermediate age, and becomes flatter in older populations. \citet{rojas2020many}, employing APOGEE data, identified a trimodal metallicity distribution in the Galactic bulge, suggesting distinct spatial distributions and kinematic structures among metal-rich, metal-intermediate, and metal-poor populations, providing important clues about their formation histories. Additionally, \citet{guiglion2024beyond} used a hybrid convolutional neural network (CNN) approach to reanalyze the Radial Velocity Spectrometer (RVS) sample and characterized the [Fe/H]-[$\alpha$/Fe] bimodality in the Galactic disk.
Further insights come from \citet{zhang2023existence}, who examined a large sample of giant stars using Gaia DR3 XP spectra. They found that a clear disc population emerges at [M/H] $\sim$ $-$1.3\,dex, with no such population present at [M/H] $< -$1.6\,dex, suggesting two distinct halo populations in the very metal-poor regime. Similarly, \citet{lian2023integrated} analyzed APOGEE metallicities and discovered that the radial [Fe/H] gradients of older stellar populations are flatter than those of younger stars. Finally, \citet{hattori2024metallicity} estimated [M/H] and [$\alpha$/M] for 48 million giant and dwarf stars in low-extinction regions using Gaia DR3 XP spectra, finding that high-[$\alpha$/M] and low-[$\alpha$/M] stars exhibit distinct kinematic properties, particularly among giants and low-temperature dwarfs.

Spectroscopic methods, while valuable, are resource-intensive in terms of telescope time and are often constrained by complex selection biases \citep[e.g.,][]{nidever2014tracing, wojno2017selection, chen2018selection}. Additionally, the faintness of the target stars limits the scope of exploration, particularly in the more distant regions of the Milky Way, such as the halo. In contrast, large-scale, high-precision, multi-band photometric surveys—such as the SkyMapper Southern Survey (SMSS; \citealt{wolf2018skymapper}), Javalambre/Southern Photometric Local Universe Survey (J/S-PLUS; \citealt{cenarro2019j, mendes2019southern}), and Gaia \citep{gaia2016gaia, gaia2018gaia, brown2021gaia}—have proven capable of reliably deriving stellar atmospheric parameters from photometry across a broad, unbiased sample of stars \citep[e.g.,][]{ivezic2008milky, yuan2015stellar, huang2022beyond, Xu2022, sun2023spar, Lu2024, Xiao2024}.

Photometric methods have shown great promise in mapping the chemical properties of the Milky Way. \citet{ivezic2008milky} demonstrated the ability to distinguish between the thin disk, thick disk, and halo components using photometrically derived metallicities from SDSS data. Building upon this, \citet{an2020blueprint, an2021blueprinta, an2021blueprintb} advanced the photometric mapping of the Milky Way by integrating spatial and chemical information with Gaia proper motions. Further, \citet{fernandez2021pristine} identified a stellar population exhibiting the rotational velocities characteristic of the thin disk, even at metallicities as low as [Fe/H] $\sim$ $-$2\,dex, using stars from the Pristine survey \citep{starkenburg2017pristine} in the direction of the Milky Way's anticenter. 
\citet{huang2019milky, huang2022beyond, huang2023beyond} applied the stellar loci fitting technique to the photometric data from SMSS DR2 and the Stellar Abundances and Galactic Evolution Survey (SAGES; \citep{fan2023stellar}) DR1. Thanks to the well-optimized narrow/medium-band $u$, $v$ filters, they were able to estimate photometric metallicities down to [Fe/H] $\sim$ $-$3.5\,dex for nearly 50 million stars. Similarly, \citet{youakim2020pristine} mapped the metallicity distribution within the Milky Way’s inner halo using Pristine Survey data, revealing a higher fraction of metal-poor stars than previously reported and estimating the likely number of metal-poor globular clusters in the halo. \citet{chiti2021stellar, chiti2021metal} explored the Milky Way’s Metallicity Distribution Function (MDF), deriving metallicities for approximately 28,000 stars down to [Fe/H] $\sim$ $-$3.75\,dex using SMSS data. Additionally, \citet{whitten2021photometric} used S-PLUS photometry to estimate effective temperatures, metallicities, carbon abundances, and carbonicity for over 700,000 stars. This data enabled them to characterize the K-dwarf halo MDF and the frequency of Carbon-Enhanced Metal-Poor (CEMP) stars in the Milky Way. Expanding on this, \citet{kim2022stars} compiled a catalog of 551,214 main-sequence stars in the local thick disk and halo, calibrated their photometric metallicities, and analyzed their kinematics, identifying structures consistent with known halo components. Finally, \citet{martin2023pristine} utilized Gaia BP/RP spectra to derive synthetic photometry of the Ca \uppercase\expandafter{\romannumeral2} H\&K region, based on narrow-band photometry from the Pristine Survey, to illustrate how the Milky Way’s structure varies with metallicity.

In this study, we leverage deep multi-band photometric data from the Kilo-Degree Survey (KiDS; \citealt{kuijken2019fourth}) and the VISTA Kilo-degree Infrared Galaxy (VIKING; \citealt{edge2013vista}) surveys to estimate photometric metallicities for approximately 820,055 stars in the Milky Way. This extensive catalog allows us to explore the Milky Way’s structure, formation, and evolutionary processes.

\section{Data} \label{data}

In this study, we aim to derive photometric metallicities using multi-band photometry, trained on a foundation of spectroscopic data.

\subsection{Photometric Data} \label{subsec:photometric_data}

The photometric data are sourced from the KiDS and VIKING surveys, which cover an area of 1350 square degrees, providing observations in both optical and near-infrared (NIR) wavelengths. The KiDS survey, conducted with the VST/OmegaCAM telescope \citep{capaccioli2011vlt, kuijken2011omegacam}, utilizes four optical filters ($u$, $g$, $r$, $i$) with a resolution of 0.2$\arcsec$/pixel. The $r$-band images are taken under optimal seeing conditions, with an average full width at half maximum (FWHM) of approximately 0.7$\arcsec$. These images reach a mean limiting AB magnitude (5$\sigma$ in a 2$\arcsec$ aperture) of 25.02 $\pm$ 0.13 mag. The remaining bands ($u$, $g$, $i$) have slightly lower resolution (FWHM $<$ 1.1$\arcsec$) and fainter limiting magnitudes of 24.23 $\pm$ 0.12 mag, 25.12 $\pm$ 0.14 mag, and 23.68 $\pm$ 0.27 mag, respectively \citep{kuijken2019fourth}. 

The VIKING survey, using the VISTA/VIRCAM instrument \citep{sutherland2015visible}, complements KiDS with five NIR bands ($Z$, $Y$, $J$, $H$, $K_{\rm S}$). The median seeing for these images is approximately 0.9$\arcsec$, with depths ranging from 21.2 mag to 23.1 mag across the passbands \citep{edge2013vista}. Together, the KiDS DR4 and VIKING DR4 releases contain photometric data for over 100 million unique sources. The deep limiting magnitudes of these datasets provide an excellent sample for investigating the Milky Way's chemical properties.

For stellar distance measurements, we used Gaia EDR3 photometric and astrometric data. Rather than directly inverting Gaia parallax measurements to estimate distances, we employed the distance catalog from \citet{bailer2021estimating}, which provides distance estimates for nearly 1.47 billion stars. This catalog is based on Gaia parallaxes and includes a weak prior that accounts for the spatial distribution of stars in the Milky Way.

Since the KiDS and VIKING surveys primarily target regions at high Galactic latitudes ($|b| > 22\degr$), the observed stars are minimally affected by extinction. The magnitudes and colors presented in this study have been corrected for reddening using $E(B-V)$ values from \citet{schlegel1998maps}, with extinction coefficients for the KiDS and VIKING passbands adopted from \citet{kuijken2019fourth}.

\subsection{Spectroscopic Data} \label{subsec:spectroscopic_data}

To derive photometric $T_{\rm eff}$ and metallicities from the multi-band photometry of the KiDS and VIKING surveys, we rely on training samples with robust spectroscopic metallicity measurements. Our primary spectroscopic dataset is drawn from the LAMOST Galactic survey. The LAMOST telescope, a quasi-meridian reflecting Schmidt instrument, is capable of simultaneously obtaining spectra for up to 4000 celestial objects within a 5$\degr$ diameter field of view. The spectra cover a wavelength range from 3700 to 9100\,$\AA$, with a resolution of $R \approx 1800$. The LAMOST DR9 release includes over 11 million spectra, comprising 242,569 galaxies, 10,907,516 stars, and 76,167 quasars. Stellar atmospheric parameters, such as metallicity ([Fe/H]), are determined using the LAMOST Stellar Parameter Pipeline (LASP; \citep{wu2014automatic}), providing reliable metallicity estimates for stars with [Fe/H] values above $-$2.5\,dex.

To ensure broader coverage across a range of stellar types and effective temperatures (3300\,K $\leq T_{\rm eff} \leq$ 8800\,K), we also incorporate data from other spectroscopic surveys, including APOGEE, SEGUE, and GALAH DR2. However, combining multiple spectroscopic datasets introduces challenges due to systematic differences in stellar parameter estimations, which can arise from variations in instrumentation and data processing methods across the surveys. To mitigate these discrepancies, we cross-match stars from LAMOST, GALAH, and SEGUE with those in APOGEE. This allows us to apply zeropoint corrections for metallicity and effective temperature, harmonizing the metallicity scales between LAMOST and the other datasets.

For the characterization of extremely metal-poor stars—a regime where many spectroscopic surveys face limitations—we augment our dataset with stars from specialized studies targeting this population \citep{li2018catalog, da2019skymapper, jacobson2015high, aguado2019pristine}. This sample includes 10,224 stars with metallicities predominantly in the range $-$3.5 to $-$2\,dex, all with [Fe/H] $< -$2\,dex. These extremely metal-poor stars enhance the precision of our metallicity estimates at the lowest end of the metallicity distribution.

\begin{table}
\bc
\begin{minipage}[]{100mm}
\caption[]{Parameter ranges for stars in the KS and KG training samples. \label{tab:effective range}}\end{minipage} \\
\setlength{\tabcolsep}{2.5pt}
\small
 \begin{tabular}{cl}
  \hline\noalign{\smallskip}
{Parameter} & {Range} \\
  \hline\noalign{\smallskip}
$T_{\rm eff}$ & [3311 -- 8846] K \\
$\log (g)$ & [0.20 -- 5.53] dex \\
$[$Fe/H$]$ & [$-$3.92 -- 0.44] dex \\
$M_{\rm g}$ & [2.26 -- 16.22] mag \\
$(g-r)_{\rm 0}$ & [$-$0.28 --1.12] mag \\
  \noalign{\smallskip}\hline
\end{tabular}
\ec
\end{table}

\subsection{Training Data Set} \label{subsec:training_set}

We constructed two distinct training sets to derive metallicities, $T_{\rm eff}$ and distances for stars. The first set, referred to as the `KS' sample, consists of objects common to both the KiDS dataset and the spectroscopic samples described earlier. The second set, named the `KG' sample, includes stars that are cross-matched between the KiDS dataset and Gaia EDR3.

The selection criteria for both the KS and KG samples are as follows:
\begin{itemize}
    \item Stars must be detected in all nine KiDS and VIKING filters, with photometric uncertainties below 0.08\,mag.
    \item The star classification probability, as indicated by the KiDS image classification (`class$\_$star'), must be at least 0.9.
    \item Reddening values, sourced from \citet{schlegel1998maps}, must satisfy $E(B-V) < 0.1$\,mag.
\end{itemize}
Additionally, for the KS sample (which includes spectroscopic data), we apply the following constraints:
\begin{itemize}
    \item The spectral signal-to-noise ratio (S/N) must exceed 30.
    \item The uncertainty in effective temperature ($\delta T_{\rm eff}$) should be less than 150\,K.
    \item Errors in surface gravity ($\delta \log(g)$) must not exceed 0.2\,dex.
    \item Metallicity errors ($\delta$[Fe/H]) should be below 0.2\,dex.
\end{itemize}
For the KG sample, which relies on Gaia data, the additional selection criteria are:
\begin{itemize}
    \item Relative errors in Gaia parallaxes ($\delta\varpi/\varpi$) must be less than 0.1.
    \item The Renormalized Unit Weight Error (RUWE) must be 1.4 or lower.
\end{itemize}

For the metal-poor star sample, we relaxed the spectroscopic data quality requirements to ensure sufficient representation of metal-poor stars in the final sample. After applying these selection criteria, the KS sample contains 11,318 stars, while the KG sample includes 136,708 stars. Table\,\ref{tab:effective range} summarizes the parameter ranges for stars within both training sets.

The metallicities in the KS sample predominantly range from $-$0.5 to $-$1.5\,dex, while the absolute magnitudes in the KG sample ($M_{\rm g}$) are primarily between 5 and 10\,mag. This distribution introduces an imbalance in the sample, which could affect the accuracy of regression models due to insufficient coverage of certain regions in parameter space. To mitigate this, we applied a grid-based averaging technique within a three-dimensional space defined by $T_{\rm eff}$, $\log(g)$, and [Fe/H] for the KS sample. We computed the average metallicity and $T_{\rm eff}$ at each grid. The grid dimensions were set to 150\,K in effective temperature, 0.05\,dex in surface gravity, and 0.02\,dex in metallicity. This process generated 9312 synthetic stars, including 8248 dwarfs and 1064 giants.

For the KG sample, we divided the parameter space into two dimensions: $M_{\rm g}$ and $(g-r)_{\rm 0}$, with grid steps of 0.01\,mag for both parameters. This resulted in 26,996 synthetic stars, with 348 giants and 26,648 dwarfs. The relatively low number of giants is due to the long exposure times of the KiDS survey and the larger relative errors in Gaia parallaxes for distant, faint stars. These gridded populations of 9,001 stars (KS sample) and 26,996 stars (KG sample) constitute our final training dataset.

\section{Methods} \label{methods}

This section describes the methodology employed in this study. First, we distinguish between giant and dwarf stars using a Random Forest Classifier (RFC) \citep{breiman2001random}. Once the stars are classified, we apply two separate Random Forest Regression (RFR) models—one for dwarfs and one for giants—to estimate photometric metallicities and $T_{\rm eff}$. Additionally, we train another pair of RFR models to predict absolute magnitudes for each stellar class.
The Random Forest (RF) algorithm \citep{breiman2001random, hastie2009elements} is an ensemble learning method that aggregates the predictions of numerous decision trees to arrive at a final estimate. Each tree in the forest is built on a randomly selected subset of features and is trained on a bootstrapped sample of the data, ensuring diversity among the trees. The final RF prediction is obtained by averaging the outputs from all trees in the ensemble. Importantly, the algorithm also calculates feature importance ($\text{Imp}(x)$) by assessing how much each feature contributes to the decision-making process across all trees. These importance scores are then averaged to provide an overall ranking of feature significance.

In this study, all RF models are implemented using the $\textit{scikit-learn}$ package \citep{pedregosa2011scikit} in Python. 
To construct the RF models, we randomly split the training data, using 70\% for model training and reserving the remaining 30\% for testing and evaluating predictive performance. In addition to the RF algorithm, we experimented with other machine learning (ML) techniques, including Support Vector Machines (SVM), Multilayer Perceptron (MLP), and Deep Learning (DL) approaches. After comparing the performance of these methods, Random Forest was found to be the most effective for accurately determining stellar parameters.

\begin{table}
\bc
\begin{minipage}[]{100mm}
\caption[]{Completeness and purity of dwarf and giant classifications in the test set. The second column shows the proportion of giants and dwarfs in the sample, the third column lists the completeness of the classifications, and the fourth column provides the purity.\label{calssify}}\end{minipage} \\
\setlength{\tabcolsep}{2.5pt}
\small
 \begin{tabular}{cccc}
  \hline\noalign{\smallskip}
{Class} & {Fraction of Sample} & {Completeness} & {Purity}\\
  \hline\noalign{\smallskip}
Dwarf & 0.96 & 0.93 & 0.99 \\
Giant & 0.04 & 0.47 & 0.31 \\
  \noalign{\smallskip}\hline
\end{tabular}
\ec
\end{table}

\begin{figure}
\centering
\includegraphics[width=0.45\textwidth]{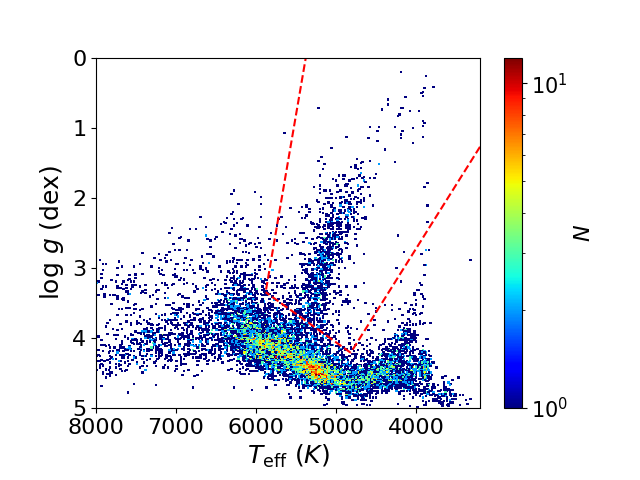}
\caption{The \textit{Kiel} diagram for the KS sample used in the dwarf-giant classification. The red polygon encloses stars classified as giants, while those outside are classified as dwarfs.}
\label{classify}
\end{figure}

\subsection{Dwarf-Giant Classification} \label{subsec:dwarf giant classification}

In this subsection, we describe the classification algorithm used to distinguish between dwarf and giant stars. We employ a Random Forest Classifier (RFC) that processes intrinsic color indices: $(u-g)_{\rm 0}$, $(g-r)_{\rm 0}$, $(r-i)_{\rm 0}$, $(i-Z)_{\rm 0}$, $(Z-Y)_{\rm 0}$, $(Y-J)_{\rm 0}$, $(J-H)_{\rm 0}$, $(H-K_{\rm s})_{\rm 0}$, along with nine dereddened magnitudes. We made an empirical classification of giants and dwarfs on the \textit{Kiel} diagram following the approach of previous works \citep[e.g.,][]{thomas2019dwarfs, guo2021three}. Fig.~\ref{classify} shows a \textit{Kiel} diagram for all stars in the KS training sample. Stars within the red polygon are classified as \textit{giants}, while those outside are considered dwarfs. The RFC provides the probability of a star being a giant, $P_{\rm giant}$, or a dwarf, $P_{\rm dwarf}$, where $P_{\rm dwarf}$ is complementary to $P_{\rm giant}$, i.e., $P_{\rm dwarf} = 1 - P_{\rm giant}$.

To evaluate the performance of our classification, we calculate the completeness and purity for both the dwarf and giant classes within the test set, as predicted by the trained RFC. These metrics are presented in Table~\ref{calssify}. The depth of the KiDS survey, which reaches beyond 25\,mag in the $g$-band, results in a sample dominated by dwarf stars. The RFC proves highly effective at classifying dwarfs, achieving a completeness of 93\% with less than 1\% contamination by giant stars. For the less frequent giants, the RFC successfully classifies nearly half, with a purity of around 31\%.

\begin{figure}
\centering
\includegraphics[width=0.32\textwidth]{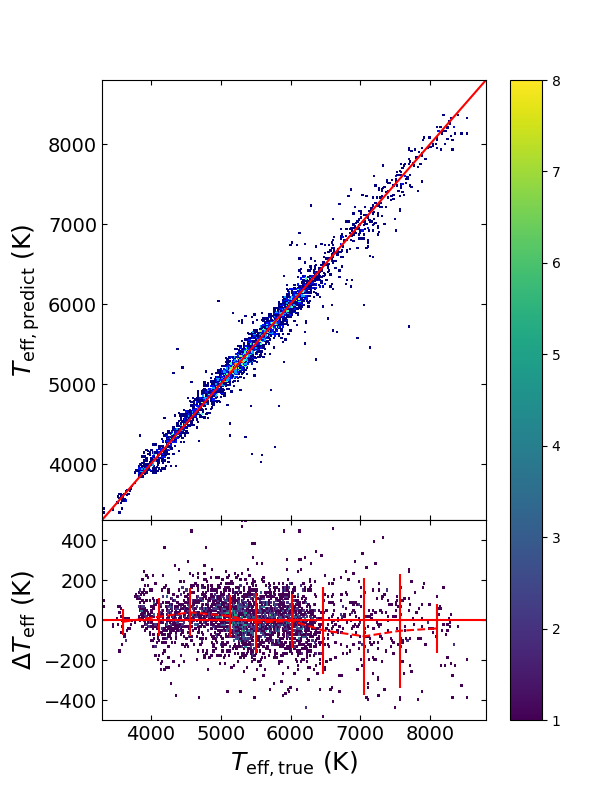}
\includegraphics[width=0.32\textwidth]{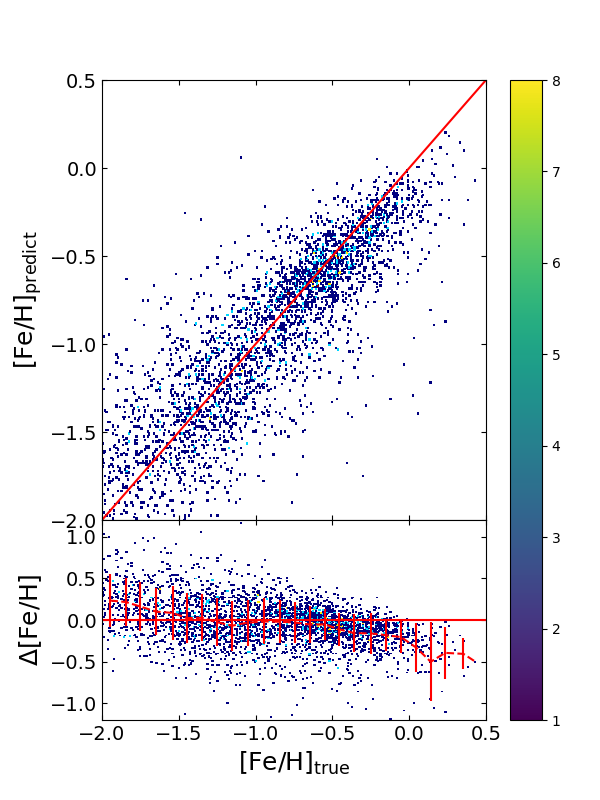}
\includegraphics[width=0.32\textwidth]{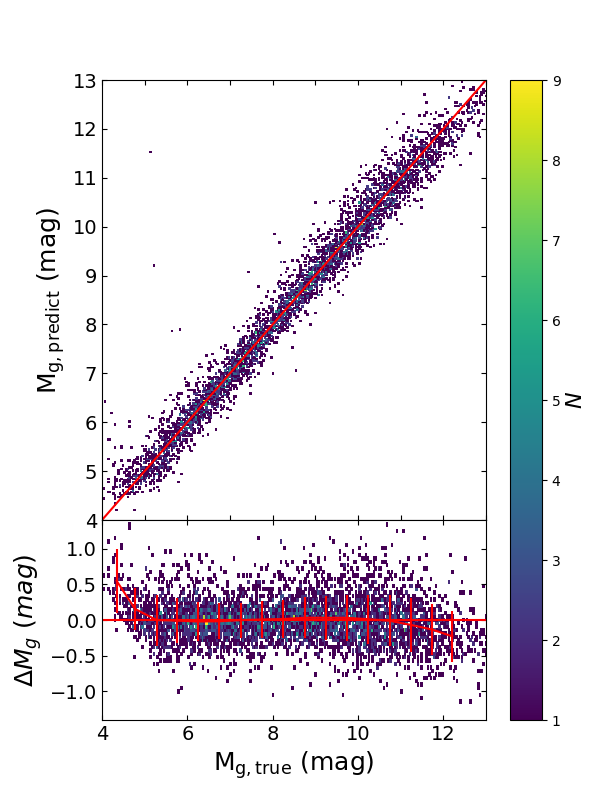}
\caption{Comparisons between true and RFR-predicted $T_{\rm eff}$ (left panel), [Fe/H] (middle panel) and $M_g$ (right panel) for the test samples. The top panels show predicted values versus true values, with the red lines indicating the one-to-one correspondence. The bottom panels show the residuals (predicted minus true values), with red lines and error bars indicating the mean differences and standard deviations within individual bins.}
\label{test_different}
\end{figure}

\subsection{Estimating Metallicities, effective temperature and absolute magnitudes} \label{subsec:metallicities and distances}

In this step, the algorithm estimates  metallicity ([Fe/H]), effective temperature ($T_{\rm eff}$) and $g$-band absolute magnitude ($M_g$) for both giant and dwarf stars. We use a RFR to predict these parameters, with eight intrinsic colors serving as input features. Once trained, the RFR models are applied to the test samples to predict [Fe/H], $T_{\rm eff}$ and $M_g$, and the results are compared to the reference values, as shown in Fig.~\ref{test_different}.

The systematic deviations between the predicted and true values of [Fe/H], $T_{\rm eff}$ and $M_g$ are minimal, with mean offsets of 0.00\,dex for [Fe/H], $-$2\,K for $T_{\rm eff}$ and $-$0.01\,mag for $M_g$. The standard deviations of the residuals are 0.34\,dex for [Fe/H], 149\,K for $T_{\rm eff}$ and 0.36\,mag for $M_g$. The dispersion of the [Fe/H] residuals is 0.28\,dex for stars with metallicity in the range $-2 <$ [Fe/H] $<$ 0.5\,dex. Across most of the metallicity range, there are no significant trends in the [Fe/H] residuals, except for the most metal-poor stars ([Fe/H] $<$ $-$2\,dex). For these stars, the algorithm tends to overestimate the metallicity compared to spectroscopic measurements, suggesting reduced reliability for stars with [Fe/H] $<$ $-$2\,dex. The broad $u$-band filter used in the KiDS survey, similar to that in SDSS, is less sensitive to changes in metallicity for stars with [Fe/H] below $-$2\,dex, which may contribute to this reduced accuracy. The residuals of the effective temperature ($T_{\rm eff}$) show no significant trends across the range of 3500\,K to 8800\,K.

For absolute magnitudes, $M_g$, most stars in the training set fall within a range of 5 to 12\,mag. The residuals show no significant trends within this range, though some deviations are observed at the extremes ($M_g < 4$\,mag and $M_g > 12$\,mag). These outliers could be due to smaller sample sizes at the boundaries of the dataset and the inherent limitations of machine learning algorithms when extrapolating beyond the core of the training distribution.

\begin{figure}
\centering
\includegraphics[width=0.32\textwidth]{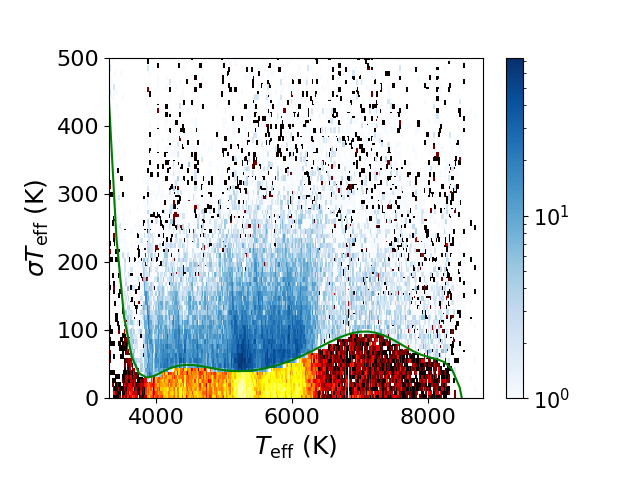}
\includegraphics[width=0.32\textwidth]{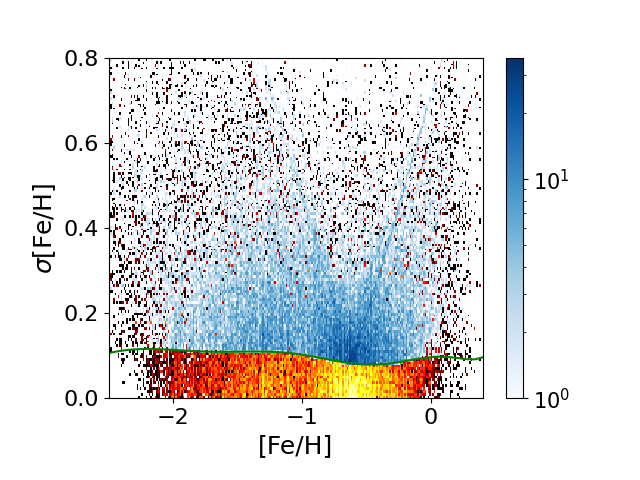}
\includegraphics[width=0.32\textwidth]{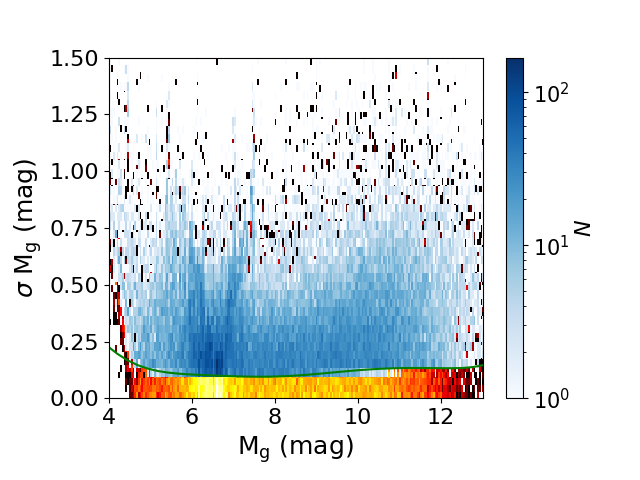}
\caption{Uncertainty estimates for $T_{\rm eff}$ (left panel), [Fe/H] (middle panel) and $M_g$ (right panel) as a function of the predicted values for the training samples. The red colormap represents the internal uncertainty, computed from the spread of predictions across the 28 RFR models. The green line shows the running dispersion derived from the training labels, representing systematic errors. The blue colormap depicts the 2D histogram of the total uncertainty, calculated as the quadratic sum of internal and systematic errors, providing a comprehensive uncertainty estimate.}
\label{error}
\end{figure}

\subsection{Estimating Uncertainties in the Training Sample} \label{subsec:determining_training_sample_uncertainties}

RF models inherently exhibit variability due to the randomization involved in bootstrap sampling and feature selection. Typically, a single partition of the dataset into training and test sets is used to train the RF models. In this study, however, we improve the robustness of our results by dividing the training sample into seven distinct training and test set pairs, each generated using a different random seed, while maintaining a consistent partition ratio of 70\% training and 30\% testing. For each of the seven splits, we train four independent RFR models, each initialized with a different random state. This approach results in a total of 28 models. The variation in random states ensures that each decision tree evaluates a different subset of features at each node, enhancing the diversity of the ensemble. These models are then used to predict metallicities ([Fe/H]), effective temperature ($T_{\rm eff}$) and absolute magnitudes ($M_g$) for the stars in our sample. The final predicted values are the averages across the ensemble of 28 models, and the standard deviations of these predictions are taken as a measure of the internal uncertainty of our method.

However, this internal uncertainty does not account for the full predictive error, as it excludes uncertainties from the input measurements. To capture the total uncertainty, we follow the approach of \citet{nepal2023gaia} and \citet{guiglion2023beyond}. Specifically, we calculate the dispersion of the differences between the true labels and those predicted by the RF models as a function of the true labels in the training sample. This dispersion provides an estimate of the precision relative to the training input, referred to as the `systematic' uncertainty.

In Fig.~\ref{error}, we present both types of uncertainties—internal and systematic—for the [Fe/H], $T_{\rm eff}$ and $M_g$ predictions. The systematic errors are measured by the spread of predictions across the 28 RFR models. For [Fe/H], the systematic errors are 0.24\,dex for giants, 0.20\,dex for dwarfs, and 0.21\,dex overall.  For $T_{\rm eff}$, the systematic errors are 90\,K for giants, 127\,K for dwarfs, and 124\,K overall. For $M_g$, the systematic errors are 0.30\,mag for giants, 0.22\,mag for dwarfs, and 0.22\,mag overall. The fitting errors, which are derived from the dispersion between the true and predicted values in the training data, are also calculated. For [Fe/H], the fitting errors are 0.14\,dex for giants, 0.12\,dex for dwarfs, and 0.12\,dex overall. For $T_{\rm eff}$, the fitting errors are 46\,K for giants, 54\,K for dwarfs, and 53\,K overall. For $M_g$, the fitting errors are 0.17\,mag for giants, 0.13\,mag for dwarfs, and 0.13\,mag overall. We combine these systematic and fitting uncertainties in quadrature to compute the final uncertainties for our predictions. As a result, the typical uncertainty in [Fe/H] is found to be 0.24\,dex, the typical uncertainty in $T_{\rm eff}$ is approximately 134\,K and the typical uncertainty in $M_g$ is approximately 0.26\,mag.

\begin{figure}
\centering
\includegraphics[width=0.55\textwidth, angle=0]{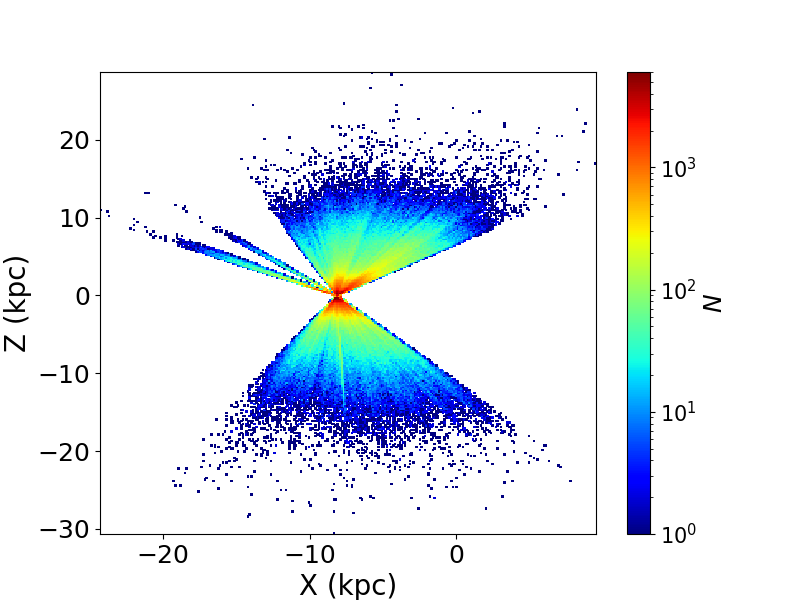}
\caption{Spatial number-density distributions for stars in the final sample. The panel shows the  $X$-$Z$ plane. The color bar at right indicates the number densities. The Sun is positioned at ($X$, $Y$, $Z$) = ($-$8.122, 0.0, 0.0)\,kpc.}
\label{X_Y_Z_distribution}
\end{figure}

\begin{table}
\bc
\begin{minipage}[]{100mm}
\caption[]{Field description for the final sample.\label{tab:description final sample}}\end{minipage} \\
\setlength{\tabcolsep}{2.5pt}
\small
 \begin{tabular}{cclc}
  \hline\noalign{\smallskip}
{Column} & {Label} & {Description} & {Unit} \\
  \hline\noalign{\smallskip}
1 & RA & Right Ascension (J2000) from KiDS DR4 & degrees \\
2 & DEC & Declination (J2000) from KiDS DR4 & degrees \\
3 & $l$ & Galactic longitude & degrees \\
4 & $b$ & Galactic latitude & degrees \\
5-8 & $u/g/r/i$ & Photometric magnitudes from KiDS DR4 & mag \\
9-13 & $z/y/J/H/K_{\rm s}$ & Photonetric magnitudes from VIKING DR4 & mag \\
14-17 & $e\_u/e\_g/e\_r/e\_i$ & Photometric uncertainties from KiDS DR4 & mag \\
18-22 & $e\_Z/e\_Y/e\_J/e\_H/e\_K_{\rm s}$ & Photometric uncertainties from VIKING DR4 & mag \\
23 & $E(B-V)$ & Reddening values & mag \\
24 & $p\_{\rm giant}$ & Classification as giant (1) or dwarf (0) & - \\
25 & Teff & Photometric effective temperature & K \\
26 & e\_Teff & Uncertainty in effective temperature estimates & K \\
27 & feh & Photometric metallicity & dex \\
28 & e\_feh & Uncertainty in metallicity estimates & dex \\
29 & $M_{\rm g}$ & Absolute magnitude in $g$-band & mag \\
30 & $e\_{M_{\rm g}}$ & Uncertainty in $M_g$ estimates & mag \\
31 & $d$ & Distance & kpc \\
32-34 & $X/Y/Z$ & Galactic Cartesian coordinates & kpc \\
35 & $R$ & Galactocentric distance & kpc \\
  \noalign{\smallskip}\hline
\end{tabular}
\ec
\end{table}

\section{Results and Discussion} \label{results}

\subsection{The Catalog} \label{final_sample}

We compiled a sample of 820,055 stars from the KiDS and VIKING surveys to estimate their metallicities, effective temperatures and $g$-band absolute magnitudes. The selection criteria were as follows:
\begin{enumerate}
    \item The stars have detections in all nine KiDS and VIKING filters with photometric uncertainties below 0.1\,mag.
    \item They are classified as stars, with the KiDS image classification parameter, `class\_star' $>$ 0.9.
\end{enumerate}
We applied the trained RFC and RFR models to the selected stars to estimate their [Fe/H], $T_{\rm eff}$ and $M_g$ values. We have identified 814,383 dwarfs and 5,672 giants in the final sample. The complete catalog, including our estimated parameters and associated uncertainties, is available online at  
\url{https://nadc.china-vo.org/res/r101527/}. Table~\ref{tab:description final sample} provides a description of the catalog contents. For each star, we list its coordinates, photometric magnitudes, and corresponding errors in each filter, along with reddening values from \citet{schlegel1998maps}. Additionally, we have included our classification of the star as a giant (1) or dwarf (0), along with our estimates for its metallicity, effective temperature, $g$-band absolute magnitudes, and the associated uncertainties. The three-dimensional (3D) positions of each star in Galactic Cartesian coordinates ($X$, $Y$, $Z$) are also provided. Here, the coordinate system is centered on the Galactic center: $X$ points away from the Sun, $Y$ aligns with the direction of Galactic rotation, and $Z$ extends towards the North Galactic Pole.

Figure~\ref{X_Y_Z_distribution} illustrates the spatial distribution of our sample stars in the $X$-$Z$ plane. The Sun is located at ($X$, $Y$, $Z$) = ($-$8.122, 0.0, 0.0)\,kpc \citep{abuter2018detection}. As shown, the majority of stars lie within 30\,kpc of the Sun, with $X$ coordinates extending from about $-$20\,kpc to 5\,kpc and $Z$ coordinates $-$30\,kpc to 30\,kpc . As previously discussed, our sample predominantly consists of dwarf stars, which are more numerous in the KiDS survey due to their fainter absolute magnitudes. Although dwarfs are detectable at relatively shallow distances, their distance estimates are accurate. However, thanks to the deep magnitude range covered by KiDS, we can probe substantial distances. For the small fraction of giant stars in our sample, the detection limit extends beyond 40\,kpc.

\begin{figure}
\centering
\includegraphics[width=0.32\textwidth, angle=0]{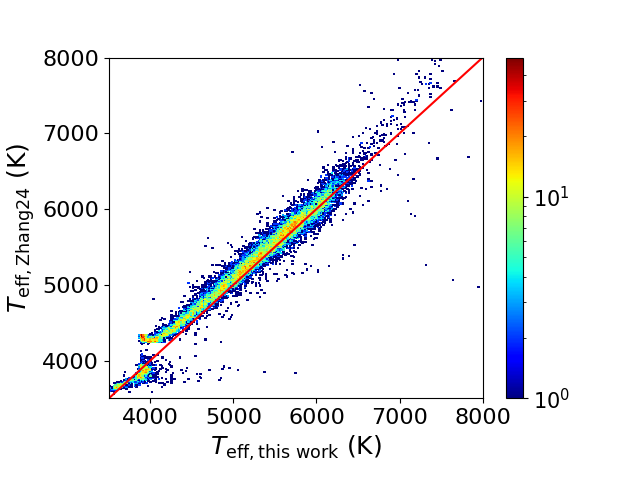}
\includegraphics[width=0.32\textwidth]{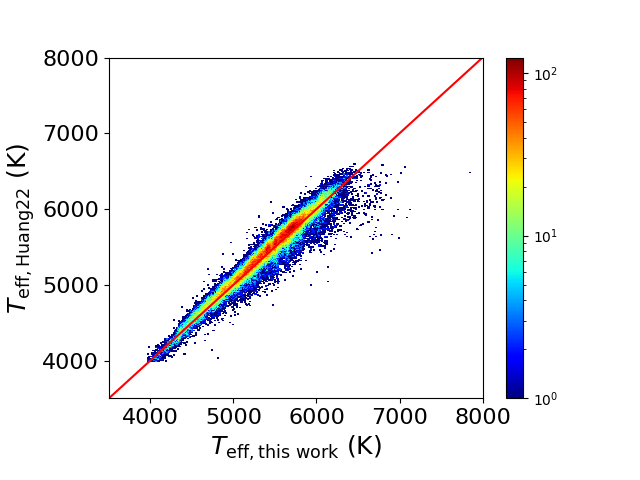}
\includegraphics[width=0.32\textwidth]{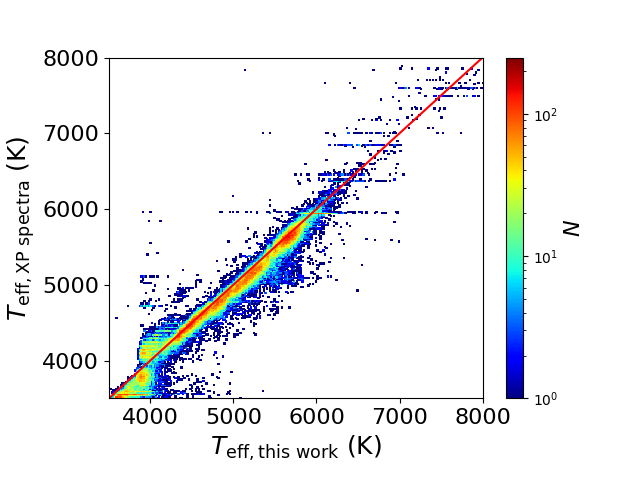}
\caption{Comparison of our effective temperature estimates with those from \citet[left panel]{zhang2024determining}, \citet[middle panel]{huang2022beyond}, and \citet[right panel]{vallenari2023gaia}. The color scales represent source density, and the red lines indicate equality between the datasets in each panel for reference.}
\label{compare_with_teff}
\end{figure}

\begin{figure}
\centering
\includegraphics[width=0.32\textwidth, angle=0]{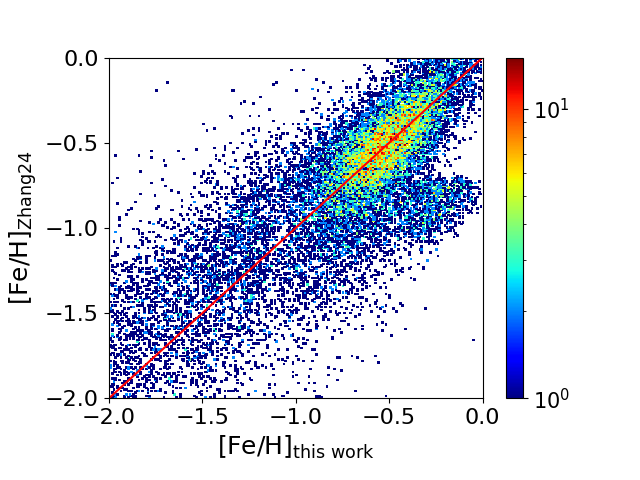}
\includegraphics[width=0.32\textwidth]{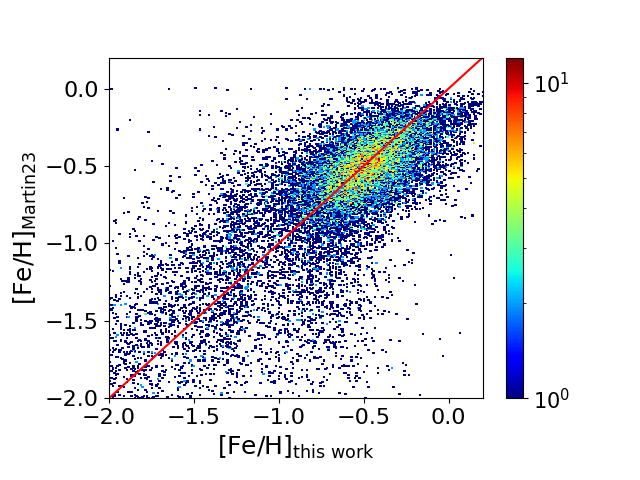}
\includegraphics[width=0.32\textwidth]{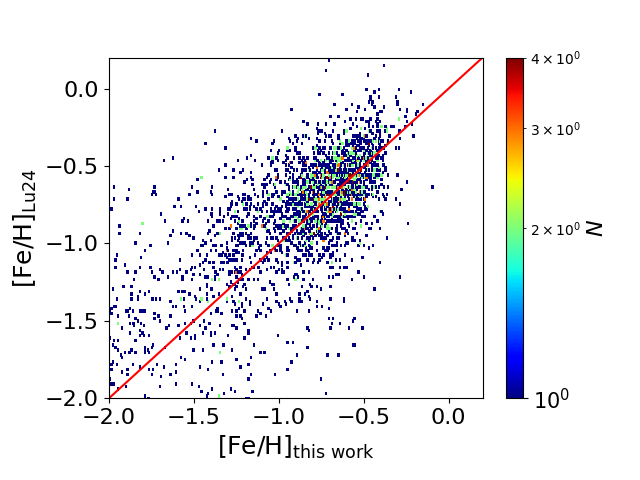}
\caption{Comparison of our metallicity estimates with those from \citet[left panel]{zhang2024determining}, \citet[middle panel]{martin2023pristine}, and \citet[right panel]{lu2024stellar}. The color scales represent the density of sources. The red lines indicate equality between the two datasets in each panel to guide the eye.}
\label{compare_with_DESI}
\end{figure}

\begin{figure}
\centering
\includegraphics[width=0.32\textwidth, angle=0]{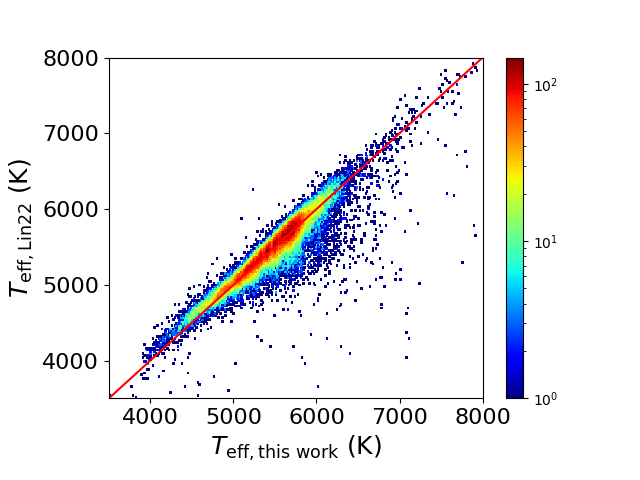}
\includegraphics[width=0.32\textwidth, angle=0]{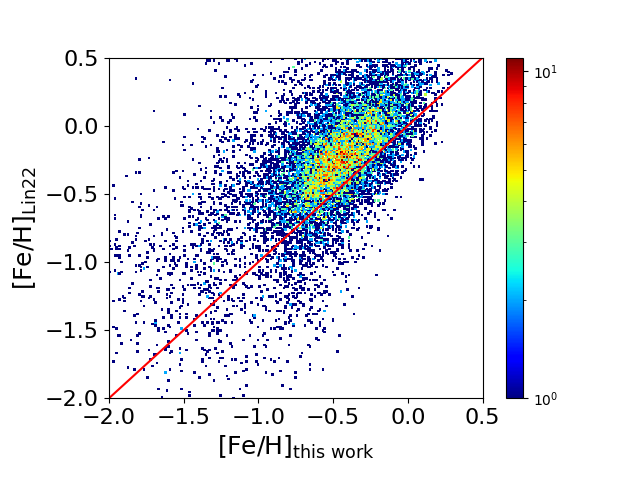}
\includegraphics[width=0.32\textwidth, angle=0]{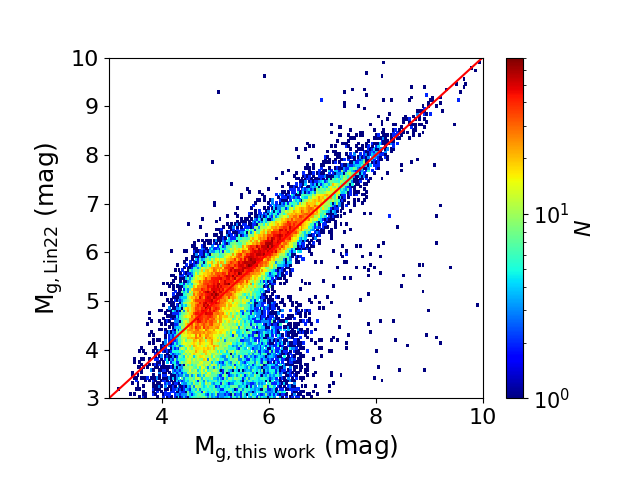}
\caption{Comparisons of our findings with those of \citet{lin2022distances} in terms of  $T_{\rm eff}$ (left panel), [Fe/H] (middle panel) and ${M_g}$ (right panel). The color scales indicate the density of sources, and the red lines represent equality between the two datasets.}
\label{compare_with_SMDR3}
\end{figure}

\begin{figure}
\centering
\includegraphics[width=0.45\textwidth]{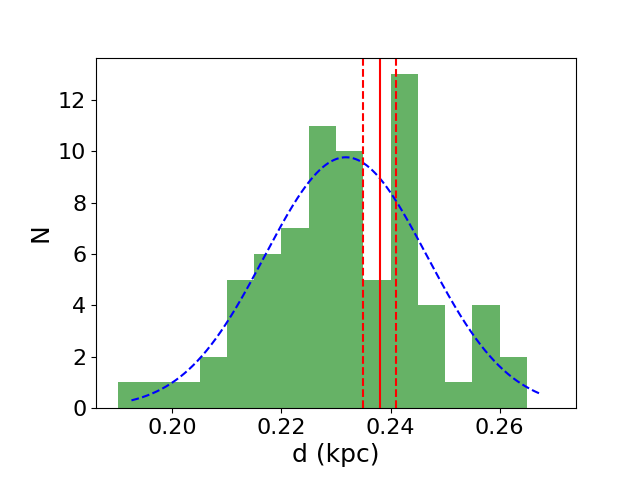}
\includegraphics[width=0.45\textwidth]{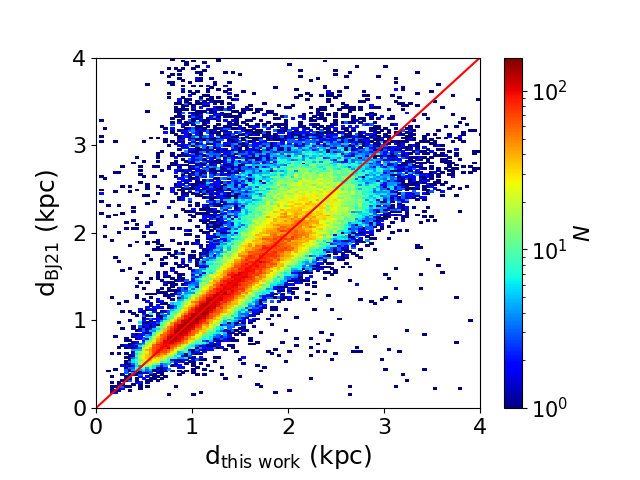}
\caption{Left panel: Histogram of our distance estimates for the member stars of the open cluster Blanco\_1, with the blue dashed line showing a Gaussian fit. The vertical red solid and dashed lines indicate the distance and uncertainty of Blanco\_1 from \citet{cantat2020painting}. Right panel: Comparison of our distance estimates with those from \citet{bailer2021estimating}. The red line denotes equality between the two datasets.}
\label{compare_with_open_cluster}
\end{figure}

\subsection{Validation of our metallicity effective temperature and distance estimates} \label{subsec:Validation of metallicity and distance estimates}

In this section, we validate the accuracy of our photometrically derived metallicities, effective temperature and distances by comparing them with independent measurements from previous studies, including \citet{lin2022distances},  \citet{huang2022beyond}, \citet{martin2023pristine}, \citet{vallenari2023gaia}, \citet{lu2024stellar}, and \citet{zhang2024determining}. 

\citet{zhang2024determining} employed the DD-PAYNE method on the low resolution spectra from Dark Energy Spectroscopic Instrument (DESI) Early Data Release (EDR) \citep{collaboration2016desi, collaboration2024early}, providing abundances for 12 elements and stellar atmospheric parameters. They obtained a precision of 0.05\,dex for metallicity and an uncertainty of 20\,K for effective temperature. We have identified 21,687 stars common to both datasets, with 143 classified as giants and 21,544 as dwarfs. Assuming the classifications by \citet{zhang2024determining} are flawless, our classification achieves a purity of 93\% for dwarfs and 67\% for giants. The left panel of Fig.~\ref{compare_with_DESI} shows a comparison between our results and their metallicities. Overall, our metallicities are in good agreement with those from \citet{zhang2024determining}, with a negligible systematic difference of $\mu = -0.02$\,dex and a standard deviation of $\sigma = 0.29$\,dex. In the left panel of Fig.~\ref{compare_with_DESI}, we observe an anomalous region where some stars have DESI metallicities between $-1.1$ and $-0.7$\,dex, while our estimates range from $-0.4$ to 0\,dex. Upon further investigation, these stars are primarily cool giants with $T_{\rm eff} \approx 4000$\,K. It is likely that the DD-PAYNE method used by \citet{zhang2024determining} underestimates the metallicities for these cooler giants. The left panel of Fig.~\ref{compare_with_teff} compares our $T_{\rm eff}$ estimates with those from \citet{zhang2024determining}. Overall, the results are consistent, with a mean difference of $\mu = 80$\,K and a standard deviation of $\sigma = 139$\,K. However, a larger systematic offset is observed for cooler stars ($T_{\rm eff}$ $<$ 4300\,K). Cross-matching with the spectroscopic data used in this study reveals an offset of 89\,K in this regime, suggesting that the \citet{zhang2024determining} results may have limitations in accurately measuring $T_{\rm eff}$ for cooler stars.

\citet{martin2023pristine} calculated metallicities by utilizing synthetic photometries obtained from the Gaia BP/RP spectra, in conjunction with the observed photometries of the narrow-band filters from the Pristine Survey. Metallicity estimates were obtained using the stellar loci method. They released two photometric metallicity catalogs: the Pristine-Gaia synthetic catalog covering most of the sky, and the higher-quality Pristine DR1 catalog, which is limited to the Pristine footprint. We compare our results with their Pristine DR1 catalog, which includes 27,565 giants and 276 dwarfs common to both datasets. The middle panel of Fig.~\ref{compare_with_DESI} shows excellent agreement between our metallicities and those from \citet{martin2023pristine}, with a negligible median difference of $\mu = -0.02$\,dex and a small standard deviation of $\sigma = 0.28$\,dex.

\citet{lu2024stellar} combined photometric data from GALEX GR6+7 AIS and Gaia. They established relationships between stellar loci, metallicity, and absolute magnitude ($M_{\rm G}$) to estimate stellar metallicities. We found 4,004 common sources between our catalog and theirs, with 20 stars consistently classified as giants and 3,825 as dwarfs. Assuming that their classifications are correct, our classification shows a purity of 87\% for dwarfs and 96\% for giants. The right panel of Fig.~\ref{compare_with_DESI} shows that our metallicities are consistent with those from \citet{lu2024stellar}, with only a small median difference of $\mu = -0.08$\,dex and a standard deviation of $\sigma = 0.31$\,dex.

\citet{lin2022distances} used a Bayesian isochrone-fitting method to derive distances, extinctions, and stellar parameters ($T_{\rm eff}$, $\log g$, [Fe/H]) for nearly 18 million stars from SMSS DR3. Their analysis incorporated 14 photometric bands from four surveys (SMSS, Gaia, 2MASS, and WISE). For the comparison, we selected stars from both catalogs with reliable measurements. In our KiDS sample, we required metallicity uncertainties of $\sigma\_[\rm Fe/H] < 0.2$ and $\sigma\_T_{\rm eff} < 150$\,K. From the \citet{lin2022distances} catalog, we selected stars with metallicity uncertainties ERR\_FEH $<$ 0.15, distance uncertainties ERR\_DIST\_ADOP/DIST\_APOD $<$ 0.1, and clean photometry (band\_flags set to `00000000000000'). This yields 13,796 common stars for the metallicity and $T_{\rm eff}$ comparison, and 57,893 stars for the distance comparison. The comparison between our results and those from \citet{lin2022distances} is shown in Fig.~\ref{compare_with_SMDR3}. For $T_{\rm eff}$, the agreement for is excellent, with a negligible mean difference of $\mu = 3$\,K and a standard deviation of $\sigma = 150$\,K. For metallicity, however, we find a significant systematic difference of $\mu = -0.24$\,dex. Given the strong agreement between our metallicities and those of \citet{zhang2024determining}, \citet{martin2023pristine}, and \citet{lu2024stellar}, we suspect that \citet{lin2022distances} may have overestimated the metallicities of their stars. For absolute magnitudes ($M_{\rm g}$), we find good agreement with \citet{lin2022distances}, especially for stars with $M_{\rm g} > 4$\,mag, where the systematic difference is only $\mu = -0.04$\,mag and the dispersion is $\sigma = 0.49$\,mag.

\citet{huang2022beyond} derived $T_{\rm eff}$ by calibrating photometric data from SMSS DR2 and Gaia EDR3 using metallicity-dependent color-$T_{\rm eff}$ relations. Among the 118,549 common sources between our catalog and theirs, we restricted the comparison to stars with errors smaller than 100\,K for reliability. The middle panel of Fig.~\ref{compare_with_teff} shows that our $T_{\rm eff}$ values agree well with \citet{huang2022beyond}, with a mean difference of $\mu = -24$\,K and a dispersion of $\sigma = 88$\,K. The observed small negative offset is primarily attributed to a systematic difference of $-68$\,K between the $T_{\rm eff}$ values reported by \citet{huang2022beyond} and the spectroscopic data used in this study.

\citet{vallenari2023gaia} derived effective temperatures ($T_{\rm eff}$) using Gaia XP spectra through the GSP-Phot module, which estimates $T_{\rm eff}$ by fitting observed XP spectra to stellar models. The right panel of Fig.~\ref{compare_with_teff} compares our $T_{\rm eff}$ estimates with those from Gaia XP spectra. For reliable comparisons, we selected 98,720 common sources from \citet{vallenari2023gaia} with well-constrained temperatures (B\_$T_{\rm eff}$\_xa $-$ b\_$T_{\rm eff}$\_x $<$ 20\,K). Our results show a mean difference of $\mu = 89$\,K and a dispersion of $\sigma = 202$\,K. While our $T_{\rm eff}$ estimates are generally consistent with those of \citet{vallenari2023gaia}, we observe a noticeable systematic offset. Additionally, discrepancies are more pronounced at lower temperatures ($T_{\rm eff}$ $\sim$ 4000\,K). A comparison between the effective temperatures from \citet{vallenari2023gaia} and spectroscopic datasets, such as those from LAMOST, reveals similar offsets and inconsistencies at the low-temperature end. 

To assess the accuracy of the distances in our sample, we compared our results with two other independent datasets: the open cluster distances from \citet{cantat2020painting} and distances derived from Gaia parallaxes by \citet{bailer2021estimating}. Due to the fact that the KiDS survey primarily covers high Galactic latitude regions and masks globular clusters \citep{kuijken2019fourth}, we identified only one open cluster suitable for comparison: Blanco\_1. Our sample contains 85 members of this cluster. As shown in the left panel of Fig.~\ref{compare_with_open_cluster}, the literature distance for Blanco\_1 is 0.237$\pm$0.003\,kpc \citep{cantat2020painting}, while our KiDS-derived mean distance is 0.225\,kpc, with a dispersion of 0.027\,kpc. The close agreement between our results and the literature values confirms the reliability of our distance estimates for these stars.

We also compared our distance estimates with those provided by \citet{bailer2021estimating}, which are based on Gaia parallax measurements. To ensure the quality of the comparison, we selected sources with reliable Gaia parallaxes, defined as having fractional parallax errors of $0.1 < \delta\varpi/\varpi < 0.2$. This selection resulted in 99,526 common sources, of which 385 are classified as giants and 99,141 as dwarfs. The comparison is shown in the right panel of Fig.~\ref{compare_with_open_cluster}. Our distances are in good agreement with those from \citet{bailer2021estimating}, with a mean difference of 0.048\,kpc and a dispersion of 0.38\,kpc.

\section{Conclusion} \label{conclusion}

Using spectroscopic training datasets, we developed RFC and RFR models to estimate the metallicity ([Fe/H]), effective temperature ($T_{\rm eff}$) and absolute magnitude ($M_{\rm g}$) of stars, based on photometric data from KiDS DR4 and VIKING DR4. The input features for the RFC model, which classifies stars as giants or dwarfs, include intrinsic colors and de-reddened magnitudes, while the output is the classification of each star as either a giant or a dwarf. The RFR models use intrinsic colors as input features to predict $T_{\rm eff}$, [Fe/H], and $M_{\rm g}$ for each star.
Our models demonstrate strong predictive performance. For $T_{\rm eff}$, the systematic difference compared to spectroscopic measurements is $-$2\,K with a scatter of 149\,K. For metallicity ([Fe/H]), the systematic difference is 0.00\,dex with a scatter of 0.28\,dex. For absolute magnitudes ($M_{\rm g}$), we find a systematic difference of $-$0.01\,mag and a dispersion of 0.36\,mag in the test sample. These results indicate the reliability of our models in deriving accurate stellar parameters from photometric data.

Using our models, we derived metallicity, $T_{\rm eff}$, and $M_{\rm g}$ values for 820,055 stars in the KiDS and VIKING datasets, including 814,383 dwarfs and 5,672 giants. When comparing our $T_{\rm eff}$ and metallicity estimates with those from spectroscopic data and other studies, we find small systematic offsets. For $T_{\rm eff}$, the offset ranges between 0 and 100\,K, with a dispersion of 100 to 200\,K. For metallicity, the systematic offset is between 0 and 0.08\,dex, with a dispersion of 0.27 to 0.37\,dex. These comparisons highlight the consistency of our results with previous studies and spectroscopic measurements. The distances derived in our final sample also show good agreement with values from the literature. Furthermore, our algorithm will be applied to larger stellar samples obtained from upcoming surveys such as the Multi-channel Photometric Survey Telescope (Mephisto) and the Chinese Space Station Telescope (CSST).

\normalem
\begin{acknowledgements}
This work is partially supported by the National Natural Science Foundation of China 12173034 and 12322304, the National Natural Science Foundation of Yunnan Province 202301AV070002 and the Xingdian talent support program of Yunnan Province. We acknowledge the science research grants from the China Manned Space Project with NO.\,CMS-CSST-2021-A09, CMS-CSST-2021-A08 and CMS-CSST-2021-B03. 

The Guoshoujing Telescope (the Large Sky Area Multi-Object Fiber Spectroscopic Telescope, LAMOST) is a National Major Scientific Project built by the Chinese Academy of Sciences. Funding for the project has been provided by the National Development and Reform Commission. LAMOST is operated and managed by the National Astronomical Observatories, Chinese Academy of Sciences.

Based on observations made with ESO Telescopes at the La Silla Paranal Observatory under programme IDs 177.A-3016, 177.A-3017, 177.A-3018 and 179.A-2004, and on data products produced by the KiDS consortium. The KiDS production team acknowledges support from: Deutsche Forschungsgemeinschaft, ERC, NOVA and NWO-M grants; Target; the University of Padova, and the University Federico II (Naples).

This publication has made use of data from the VIKING survey from VISTA at the ESO Paranal Observatory, programme ID 179.A-2004. Data processing has been contributed by the VISTA Data Flow System at CASU, Cambridge and WFAU, Edinburgh.

This work presents results from the European Space Agency (ESA) space mission Gaia. Gaia data are being processed by the Gaia Data Processing and Analysis Consortium (DPAC). Funding for the DPAC is provided by national institutions, in particular the institutions participating in the Gaia MultiLateral Agreement (MLA). The Gaia mission website is \url{https://www.cosmos.esa.int/gaia}. The Gaia archive website is \url{https://archives.esac.esa.int/gaia}.

Funding for the Sloan Digital Sky Survey IV has been provided by the Alfred P. Sloan Foundation, the U.S. Department of Energy Office of Science, and the Participating Institutions. SDSS acknowledges support and resources from the Center for High-Performance Computing at the University of Utah. The SDSS web site is \url{www.sdss4.org}.

This work made use of the Third Data Release of the GALAH Survey (Buder et al. 2021). The GALAH Survey is based on data acquired through the Australian Astronomical Observatory, under programs: A/2013B/13 (The GALAH pilot survey); A/2014A/25, A/2015A/19, A2017A/18 (The GALAH survey phase 1); A2018A/18 (Open clusters with HERMES); A2019A/1 (Hierarchical star formation in Ori OB1); A2019A/15 (The GALAH survey phase 2); A/2015B/19, A/2016A/22, A/2016B/10, A/2017B/16, A/2018B/15 (The HERMES-TESS program); and A/2015A/3, A/2015B/1, A/2015B/19, A/2016A/22, A/2016B/12, A/2017A/14 (The HERMES K2-follow-up program). We acknowledge the traditional owners of the land on which the AAT stands, the Gamilaraay people, and pay our respects to elders past and present. This paper includes data that has been provided by AAO Data Central (datacentral.org.au).
\end{acknowledgements}
  
\bibliographystyle{raa}
\bibliography{kidsmetallicity}

\end{document}